# Single Crystal Functional Oxides on Silicon

Saidur Rahman Bakaul[1], Claudy Rayan Serrao[1], Michelle Lee[2], Chun Wing Yeung, Asis Sarker[1], Shang-Lin Hsu[4], Ajay Yadav[3], Liv Dedon[3], Long You[1], Asif Islam Khan[1], James David Clarkson[3], Chenming Hu[1], Ramamoorthy Ramesh[2,3,4], Sayeef Salahuddin[1,4*]

1 Dept. of Electrical Engineering and Computer Sciences, University of California, Berkeley
2 Dept. of Physics, University of California, Berkeley
3 Dept. of Material Science and Engineering, University of California, Berkeley
4 Material Science Division, Lawrence Berkeley National Laboratory, Berkeley,

*To whom correspondence should be addressed; E-mail: sayeef@berkeley.edu

**Single crystalline thin films of complex oxides show a rich variety of functional properties such as ferroelectricity, piezoelectricity, ferro and antiferromagnetism etc. that have the potential for completely new electronic applications (*1-2*). Direct synthesis of such oxides on Si remains challenging due to the fundamental crystal chemistry and mechanical incompatibility of dissimilar interfaces (*3-16*). Here we report integration of thin (down to 1 unit cell) single crystalline, complex oxide films onto Si substrates, by epitaxial transfer at room temperature. In a field effect transistor using a transferred $Pb_{0.2}Zr_{0.8}TiO_3$ (PZT) layer as the gate insulator, we demonstrate direct reversible control of the semiconductor channel charge with polarization state. These results represent the realization of long pursued but yet to be demonstrated single crystal functional oxides on-demand on silicon.**

A significant number of single crystalline complex oxides show ferroic order and a variety of correlated phenomena (*1,2*). Consequently, extensive research effort is currently ongoing in the investigation of these materials both for fundamental science and potential applications. For many of the novel functionalities, it is important to retain the single crystal nature of these oxides when they are finally interfaced with Si electronics. In addition, it has been long postulated that integration of single crystal functional oxides with silicon could resolve some of the most critical problems in existing applications such as the memory retention time in Ferroelectric Random Access Memory (*3*). As a consequence, there is currently a significant effort to integrate functional complex oxides on Silicon (*4-17*). However, owing to large difference in interfacial chemistry and the typically high temperatures and oxidizing environments needed for the growth of such oxides, direct epitaxial synthesis on Si continues to pose a significant synthesis challenge (*6-10*). Such integration is mostly achieved by growing an appropriate buffer layer (*9, 11-16*), which then acts as a template for synthesis of subsequent layers either by epitaxy or other techniques. Synthesis of a ferroelectric without a buffer layer has also been demonstrated (*17*). However, a common problem in all these methods comes from the electronic incompatibility of the interfaces that leads to dangling bonds and trap states. These trap states in turn dominates the electronic behavior and decouples the functional oxides from the underlying Si channel. For example, despite the pioneering work of epitaxial growth of a ferroelectric layer on silicon without a buffer layer in (*17*), a direct and reversible control of the Si channel charge could not be achieved.

In this paper, we present a fundamental advancement in the integration of such dissimilar materials. This is achieved by epitaxial transfer of single crystalline functional oxides directly onto Si. Due to the fact that the process can be carried out at room temperature, it avoids the interface chemistry and thermal issues described above. We demonstrate transfer of functional oxides as thin as 1 unit cell (4 Å), which is almost three orders of magnitude thinner than any other transfer technique reported for complex oxides. The lattice structure, surface morphology, piezoelectric coefficient, dielectric constant, polarization switching and spontaneous and remnant polarization of the transferred ferroelectric oxide are commensurate with those of the as-grown films on lattice matched oxide substrates. Remarkably, when a transferred PZT is used as the gate of a silicon-on-insulator (SOI) transistor, it shows clear control of the channel charge with ferroelectric polarization evidenced in the signature anti-clockwise hysteresis loop and an abrupt jump in the current, attesting to high quality interface and single crystalline nature of the transferred film respectively. We also demonstrate transfer of single crystalline superlattices and multiferroic heterostructures on Si that illustrate the tremendous flexibility offered by the technique reported in this work.

For epitaxial transfer, we start by growing single crystal, 0.4-100 nm thick PZT on 20 nm thick $La_{0.7}Sr_{0.3}MnO_3$ (LSMO) coated $SrTiO_3$ (STO) substrate by using pulsed laser deposition (PLD). Subsequently, the LSMO layer is wet etched. This releases the layer(s) sitting above it (Fig. 1A), which is then carried by a transfer stamp based on a poly methyl methacrylate (PMMA) and placed on the target substrate such as Si. High resolution transmission electron microscopy (HRTEM) reveals atomically sharp interfaces and no interfacial layer when Si surface is properly passivated (Fig. 1B and fig. S3A to S3B). Similar results are obtained when stack with multiferroic ($SrRuO_3/BiFeO_3/CoFeB/MgO$) and superlattices ($CaTiO_3/SrTiO_3)_6$ are transferred (Fig. 1C and 1D). Fig. 2 shows the structural characteristics of transferred films of PZT on Si. The root mean square (RMS) roughness of the transferred PZT is 0.61 nm (Fig. 2A) which is comparable to that of the as-grown film (0.42 nm; fig. S1A). The bottom surface of the PZT, which was released from LSMO, shows a RMS roughness of 0.67 nm (Fig. 2B). This indicates that the surface morphology of PZT is insensitive to the etch chemistry and removal of LSMO. The θ-2θ scan of the transferred film using X-ray Diffraction (Fig. 2D) is essentially identical to the as-grown film (Fig. 2C) and shows peaks only from the PZT (001) and Si (00l) family planes, suggesting that the transferred PZT is a single crystal. The lattice constants for the as-grown and the transferred PZT are 4.14 Å and 4.15 Å respectively and the full width half maxima measured from the rocking curves are 0.54° and 0.53°. This suggests that the overall film quality remains intact after the transfer process. Similar behavior is observed when PZT is transferred on other surfaces such as 5-nm amorphous $Al_2O_3$ coated Si, thermally grown amorphous $SiO_2$ coated Si, sputter deposited amorphous Au coated Si, single oxide substrates such as LSMO on STO etc. (fig. S3).

Next we studied the electromechanical properties of the transferred PZT using the piezoelectric force microscopy (PFM). As shown in Fig. 3A, the ferroelectric domains of the transferred PZT on Si could be reversibly poled by applying oppositely directed electric fields from the PFM tip. The domains thus obtained retained their respective polarization states even after 24 hours. Fig. 3B shows the $d_{33}$-V loop for the transferred PZT on Si. The $d_{33}$ amplitude is similar to that obtained in the as-grown film (fig. S5). Next we explore the Polarization (P)-field (E) and Capacitance (C)-E characteristics. Fig. 3C and 3D show the results for the case where an epitaxial tri-layer $SrRuO_3$(SRO)/PZT/SRO heterostructure on LSMO buffered STO substrate was grown and subsequently transferred onto a Si substrate. The saturation polarization (~75 μC/cm$^2$) and the peak capacitance (~1.6 μC/cm$^2$) are similar to a typical as-grown film. The hysteresis is symmetric with the V=0 point because of a symmetric boundary condition on top and bottom for the PZT film (*18*). Importantly, the results in Fig. 3C and 3D demonstrate that the transfer method works equally well for multiple layers and therefore any arbitrary heterostructure can be transferred in this way. Monitoring the voltage across the ferroelectric after application of a pulsed voltage shows a transient decrease with time, characteristic of the intrinsic polarization switching (see supplementary section 7 for details) (*19-21*).

To check the electronic quality of the interface, we demonstrate a functional Si Field Effect Transistor (FET) with a transferred PZT layer as the gate oxide. We exploit one of the major strengths of the transfer process, namely, a single crystalline ferroelectric can be transferred onto any arbitrary surface, such as Si/SiO$_2$ (3nm) surface. The Si/SiO$_2$ interface ensures excellent surface for the channel and at the same time provides a large band-offset with the channel that stops hot electrons from easily tunneling into the ferroelectric atop it. The PZT is then transferred onto the channel to form the gate.

Fig. 4A shows the normalized, frequency-dependent capacitance of a Si/SiO$_2$ capacitor with and without the transferred PZT on top. The dispersion is identical for both, indicating that the transfer of PZT does not degrade the quality of the interface. The impedance angle is close to 90° for both capacitors over the entire frequency range. Similar behavior is seen for Si/Al$_2$O$_3$ interfaces (fig. S9). Fig. 4B shows the schematic of the fabricated transistor. Fig. 4C shows the $logI_D$-$V_G$ characteristics. There are two important points about the $logI_d$-$V_g$ characteristic. Firstly, the $logI_d$-$V_g$ shows counterclockwise hysteresis for the n-type transistor which is a characteristic signature of the ferroelectric control of the charge. Secondly, the abrupt jump in the current indicates that the ferroelectric PZT switches abruptly as expected in a single crystalline structure. The handedness of the hysteresis and the abruptness in the current together demonstrate the successful integration of a functional, single crystalline oxide onto a Si device, a goal that has been long pursued but has so far been elusive (*17*). All of the $I_D$-$V_G$ loops are repeatable (fig. s12).

Our work is a fundamental advancement over prior transfer methods that have been explored before for ferroelectrics (such as the smart-cut techniques where only microns thick films have been transferred and a typical surface RMS roughness of 11-70 nm is observed (*22-25*) due to ion damage.. By contrast, we have integrated films with thickness much smaller than this roughness ranges down to a single unit cell. The generality of our approach paves the way integrate complex oxides on not only Si but also other semiconductors such as GaN where the polarization of a single crystalline ferroelectric could be used to counteract the built-in polarization. Epitaxially transferred semiconductors is a commercial technology (*27*). This indicates that the reported technique should be scalable to commercially relevant sizes, thereby enabling many novel applications in electronics and multiferroic spintronics (*26, 28-31*).

**Acknowledgement:**
This work was supported in part by the ONR, ARO YIP award, the AFOSR YIP award, the STARNET LEAST center, the NSF E$^3$S center and the IRICE program at Berkeley. The authors gratefully acknowledge discussion with Dr. Guneeta Singh Bhalla who first brought our attention to wet etching of manganite films. All additional data are available in the supplementary materials.

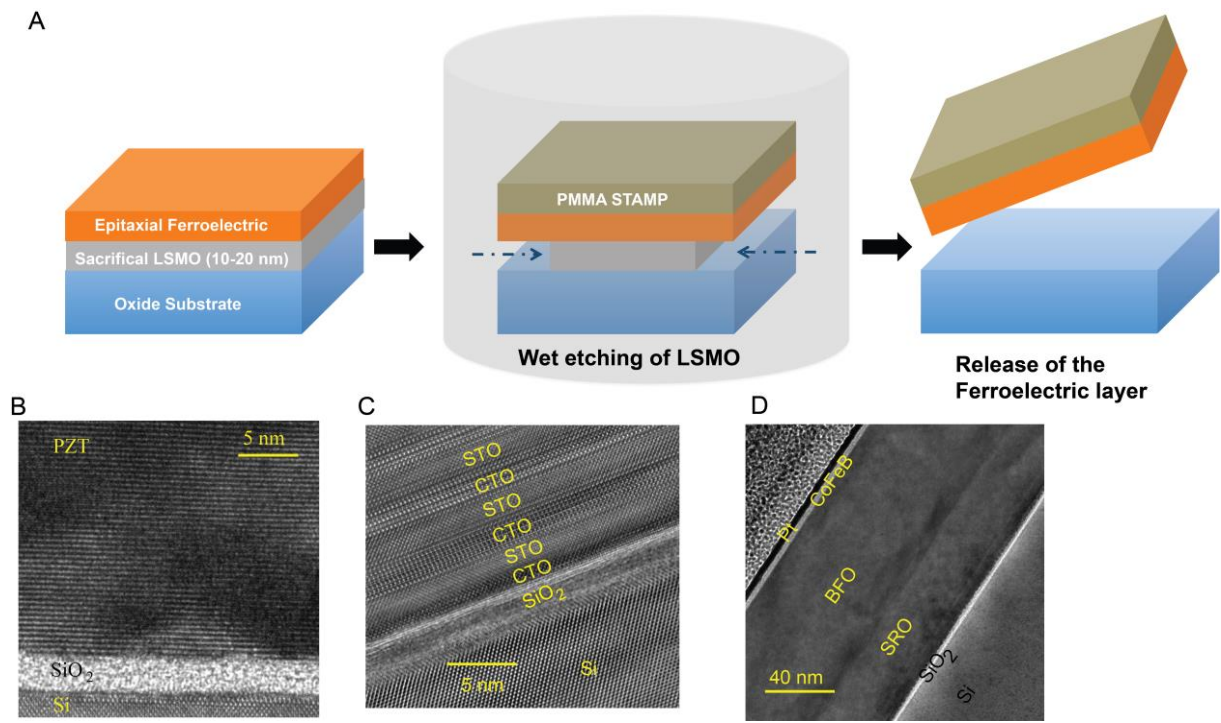

**Fig. 1. Epitaxial ferroelectric films on Si. (A) Transfer process.** Epitaxial thin films (1 unit cell -100 nm) of ferroelectric oxides are grown on lattice matched substrates with a thin (10-20 nm) sacrificial layer using PLD method. The stack is then immersed in a diluted KI+HCl solution, which isotropically etches LSMO. A PMMA handle is used to transfer the released ferroelectric layers onto Si and other substrates. TEM images of the transferred (**B**) PZT, (**C**) (CTO/STO)$_6$ superlattices and (**D**) SRO/BFO/CoFeB/Pt multilayers on Si substrate.

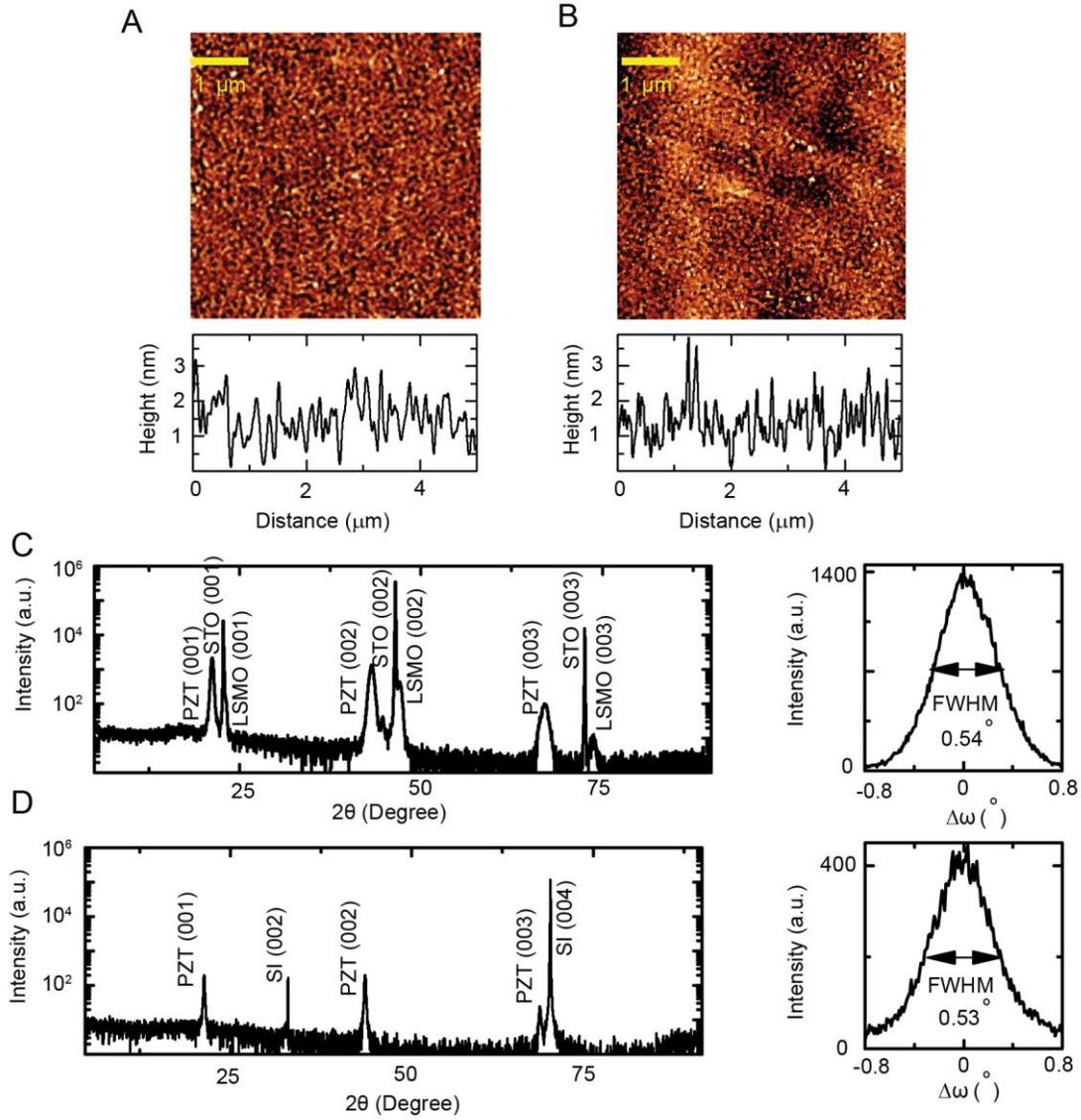

**Fig. 2. Structural characterization of the as-grown PZT and the transferred PZT on Si. A,B,** AFM images of the top and bottom surfaces of transferred PZT. The top surface is probed when PZT is sitting on Si and the bottom surface is probed by placing PZT/PMMA bilayer inverted on Si. The RMS roughness of top and bottom surfaces is 0.61 and 0.67 nm, respectively. These are comparable to 0.41 nm roughness of the source PZT film's top surface (fig. S1). **C,D** θ-2θ scan and rocking curve around PZT (002) reflection peak of the source PZT on STO/LSMO substrate and transferred PZT on Si (001). The absence of any phase other than the 001 family of planes of Si and PZT points that the transferred PZT is single crystalline. The FWHM of the source and transferred PZT's rocking curves are also very similar.

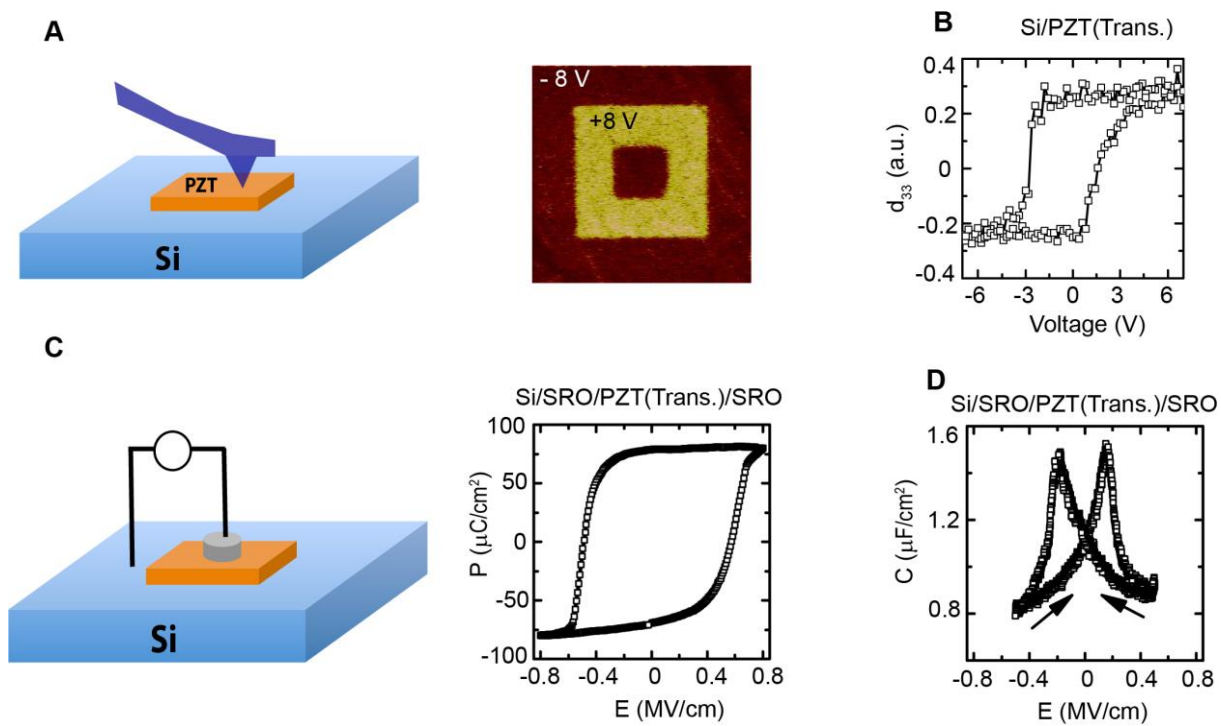

**Fig. 3. Piezoelectric and ferroelectric properties of the transferred PZT on Si. A,** Piezoforce microscopy of the transferred layer. The ferroelectric domains can be reversibly poled and the states are very stable. **B,** the $d_{33}$ coefficient of the transferred PZT on Si. **C, D,** P-E and C-E loop of a SRO/PZT/SRO transferred on highly doped Si substrate.

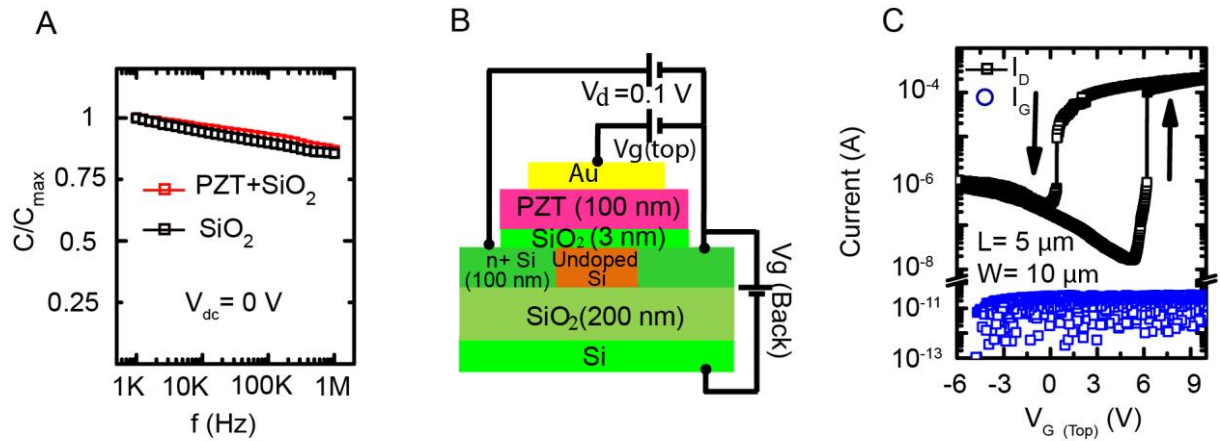

**Fig. 4. Single crystal PZT gated Si channel transistor. A,** Frequency dependent capacitance of Si/SiO$_2$ and Si/SiO$_2$/transferred PZT. The capacitor size is 22×22 μm$^2$. **B,** Cross-sectional schematic diagram of the fabricated SOI transistor. The length and width of the Si channel region are 5 μm and 10 μm, respectively, whereas gate electrode length is 20 μm. **C,** I$_D$-V$_G$ (top gate) characteristics of the ferroelectric PZT gated transistor at V$_G$ (back gate)=0. The counter-clockwise hysteresis and two order of abrupt current change in the I$_D$-V$_G$ characteristics demonstrates the control of the channel charge by the polarization of the transferred PZT layer.

**References:**

1. M. Dawber, K. M. Rabe, J. F. Scott, *Rev. Mod. Phys.* **77**, 1083-1130 (2005).
2. I. Masatoshi, A. Fujimori, Y. Tokura, *Rev. Mod. Phys.* **70**, 1039-1263 (1998).
3. T. P. Ma and J. P. Han, *IEEE Elect. Dev. Lett.* **23**, 386 (2002).
4. E. Defaÿ, *Ferroelectric dielectrics integrated on silicon* (Wiley, 2011).
5. Y. Shichi *et al. Jpn. J. Appl. Phys.* **33**, 5172-5177 (1994).
6. R. A. McKee, F. J. Walker, M. F. Chisholm, *Phys. Rev. Lett.* **81**, 3014 (1998).
7. R. Ramesh, *et al.*, *Appl. Phys. Lett.* **63**, 3592 (1993).
8. J. H. Haeni *et al., Nature* **430**, 758-761 (2004).
9. S. H. Baek *et al.*, *Science* **334**, 958-961 (2011).
10. C. Dubourdieu *et al.*, *Nature. Nanotech.* **8**, 748-754 (2013).
11. A. Chin, M. Y. Yang, C. L. Sun, S. Y. Chen, *IEEE Elect. Dev. Lett*. **22**, 336-338 (2001).
12. H. Takasu, *Microelectronic Engineering* **59**, 237-246 (2001).
13. S. Sakai, M. Takahashi, *Materials* **3**, 4950-4964 (2010).
14. E. Tokumitsu, K. Okamoto, H. Ishiwara, *Jpn. J. Appl. Phys.* **40**, 2917-2922 (2001).
15. Y. Wang, *et al.*, *Appl. Phys. Lett*. **80**, 97 (2002).
16. T. Hirai, K. Teramoto, T. Nishi, T. Goto, Y. Tarui, *Jpn. J. Appl. Phys.* **33**, 5219-5222 (1994).
17. M. P. Warusawithana *et al., Science* **324**, 367-370 (2009).
18. C. B. Eom, *et al. Science* **258** 1766-1769 (1992).
19. A. I. Khan *et al. Nature Mat.* **14**, 182-186 (2015).

20. S. Salahuddin, S. Datta, *Nano Lett*. **8**, 405-410 (2008).
21. A. I. Khan *et al., Applied Physics Letters* **99,**113501 (2011).
22. M. Levy *et al.*, *Appl. Phys. Lett.* **73**, 2293 (1998).
23. T. Izuhara et al., *Appl. Phys. Lett.* **82**, 616 (2003).
24. M. Alexe, U. Gösele, *Wafer Bonding Applications and Technology* (Springer, 2004).
25. P. Young-Bae, M. Bumki, K. J. Vahala, H. A. Atwater, *Adv. Mater*. **18**, 1533–1536 (2006).
26. Y. Qi, *et al., Nano Lett.***11**, 1331-1336 (2011).
27. B. M. Kayes *et al.*, *Photovoltaic Specialists Conference (PVSC), 2011 37th IEEE*. (2011).
28. V. V. Zhirnov, R. K. Cavin, *Nature Nanotech*. **3**, 77–78 (2008).
29. L. Li *et al.*, *Science* **332,** 825-828 (2011).
30. J. T. Heron *et al.*, *Nature* **516,** 370-373 (2014).
31. P. Aguado-Puente *et.al.*, *Phys. Rev. Lett.* **107,** 217601 (2011).

# Supplementary Materials for
# Single Crystal Functional Oxides on Silicon

Saidur Rahman Bakaul[1], Claudy Serrao[1], Michelle Lee[2], Chun Wing Yeung, Asis Sarker, Shang-Lin Hsu[4], Ajay Yadav[3], Liv Dedon[3], Long You[1], Asif Islam Khan[1], James David Clarkson[4], Chenming Hu[1], Ramamoorthy Ramesh[2,3,4], Sayeef Salahuddin[1,4*]

1 Dept. of Electrical Engineering and Computer Sciences, University of California, Berkeley
2 Dept. of Physics, University of California, Berkeley
3 Dept. of Material Science and Engineering, University of California, Berkeley
4 Material Science Division, Lawrence Berkeley National Laboratory, Berkeley,

*To whom correspondence should be addressed; E-mail: sayeef@berkeley.edu

This PDF file includes:

Materials and Methods

Supplementary Text

Figs. S1 to S12

References

**Materials and Methods:**

**Materials growth by pulsed laser deposition (PLD)**

STO/LSMO20/PZT:

20 nm of LSMO is grown on the STO (001) substrate at 750 °C with repetition rate of 2 Hz. The film is cooled down to 630°C at a rate of 5°C /min. PZT is grown at this temperature using 10 HZ repetition rate. Both the layers are grown at the oxygen background pressure of 100 mTorr and laser energy density of 1 J/cm$^2$. The film is cooled down to room temperature at the rate of 5°C /min in 1 atm pressure of oxygen. Figure S1a and S1b shows the surface morphology of PLD grown STO/LSMO20/PZT100 sample, which is used as the source substrate for the transferred PZT flakes shown in main text Fig. 2.

STO/LSMO20/SRO15/PZT60/SRO15:

After growing 20 nm of LSMO at 750 °C on the STO (001) substrate, the film is cooled down to 700 °C (cooling rate 5 °C/min) where 15 nm of SRO is grown at 10 HZ. Then the film is cooled down to 630 °C where 60 nm of PZT is grown. After the PZT growth 15 nm of SRO is grown at same temperature. All the layers are grown at the oxygen background pressure of 100 mTorr and laser energy density of 1 J/cm$^2$. The film is cooled down to room temperature at the rate of 5 °C/min in 1 atm pressure of oxygen.

DSO/LSMO20/BTO:

20 nm LSMO is grown following the same recipe as described earlier. Then the film is cooled down to 600 °C at a rate of 10 °C/min and BTO is grown at that temperature. The oxygen background pressure is 20 mTorr, laser energy density is 1.5 J/cm$^2$ and repetition rate is 10 Hz. The film is cooled down to room temperature at the rate of 10 °C /min in 1 atm pressure of oxygen.

Superlattices:

Superlattices of SrTiO$_3$/CaTiO$_3$, e.g. ((STO)6 /(CTO)1)6, were synthesized using Reflection High Energy Electron Diffraction (RHEED) - assisted PLD. To ensure stoichiometric transfer of STO and CTO, growth temperature was set at 700 C and growth pressure was 50 mTorr with both targets being ablated by a laser fluence of 1.5 J/cm$^2$. The growth was monitored using RHEED with Frank-van der Merwe layer-by-layer growth mode present throughout the process.

DSO/SRO40/BFO70/CoFeB4/Pt4:

These films were prepared on single-crystalline (110) DyScO$_3$ substrates by PLD. For SRO and BFO films substrate temperatures were 690 °C and 700 °C and Oxygen pressures were 50 mTorr and 100 mTor , respectively. The films were grown at a repetition rate of 8 Hz with a laser fluence of 1.1 J.cm$^{-2}$. After growth, the samples were cooled to room temperature in an Oxygen pressure of 750 Torr.

**Details of single crystal FE transfer method**

The key to transfer single crystal FE onto Si is using a suitable sacrificial layer which, a) allows the crystalline growth of the FE layer on top of it and b) can be selectively etched away without affecting the ferroelectric layer. Fig. 1 shows the steps involved in the transfer process. We grow single crystal PZT on 20 nm thick La$_{0.7}$Sr$_{0.3}$MnO$_3$ (LSMO) coated STO substrate by using PLD. Then poly methyl methacrylate (PMMA) layer is spin coated, which serves as the transfer stamp. We chose PMMA over other typical stamps such as PDMS, since the adhesion between PZT and PMMA is better than that between PZT and PDMS. Moreover, PMMA can be easily and cleanly removed by organic chemical such as acetone which helps to get better yield in transferring nano materials (*1*). PMMA 950A4 is spin coated at 4000 rpm for 30 seconds, followed by a baking at 120 °C for 1 minute, which leaves a uniform film of 200-250 nm thickness. The second and the most important task to successfully transfer single crystal PZT is finding an etchant which removes only the sacrificial layer LSMO and does not react with the PMMA stamp and PZT. We use KI (4 mg) + HCl (5 mL) + H$_2$O (200 mL) for etching (*2*). Since LSMO/PZT stack is covered by PMMA stamp, LSMO is slowly etched away only from the side. The possible chemical reaction between LSMO and the etchant is:

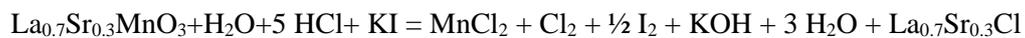

$$La_{0.7}Sr_{0.3}MnO_3 + H_2O + 5\ HCl + KI = MnCl_2 + Cl_2 + \tfrac{1}{2} I_2 + KOH + 3\ H_2O + La_{0.7}Sr_{0.3}Cl$$

After 12 hours, PZT/PMMA stack is released and collected from the solution. Then we wash it by DI water and dry up by N$_2$ gas. The stamp is transferred by using micromanipulator and a gentle force on the target substrate, followed by removal of PMMA by acetone. Following this procedure we have transferred various types of ferroelectric materials and heterostructures such as PZT, BTO, CTO/STO superlattices, SRO/BiFeO$_3$/CoFeB/Pt and SRO/PZT/SRO. The transfer process works equally well for all of these structures, including 1 unit cell thick PZT. Fig. S2A and S2B show the topography images of the transferred 1 unit cell thick PZT on Si. The height profile clearly shows a step of approximately 0.435 nm between Si and PZT flake, where the error bar is ±0.1 nm. This step height matches with the estimated thickness found from the growth rate of the source film, confirming the unit cell transfer capability of our technique. Fig. S2C and S2D show topography images of the transferred BTO on Si. The XRD ɵ-2ɵ scan shown in Fig. S2E and S2F clearly demonstrate the single phase of BTO before and after transfer process.

**Supplementary Text:**
**1. TEM images of transferred oxide layers**
Fig. S3 A-C show the TEM images of the transferred PZT on 5nm Al$_2$O$_3$ coated Si, 10 nm Au coated Si and transferred SRO/PZT/SRO heterostructures on Si/SiO$_2$, respectively. The images reveal atomically sharp interfaces. The SiO$_2$ layer is necessary for transistor application and could be easily avoided by performing the transfer in an inert environment. The area near top surface of PZT also reveals no microstructural damage, as shown in Fig. S4B. All these TEM images points that wide range of materials including multilayers can be placed on arbitrary substrate in pristine condition.

## 2. Extracting lattice constant of the bottom and top portion of transferred PZT on Si/SiO$_2$ by Fourier Transformation of the TEM image

Fig. S4A, B show the TEM images of 32×32 nm area of transferred PZT near the bottom and top surface, respectively. The spacing between the periodic lines seen in the TEM image, stemming from the repetitive arrangement of PZT planes is a direct measure of the c-axis lattice parameter of PZT. To determine the exact periodicity from the images we perform Fourier transformation on them and the resultant color coded images are shown in the insets of figure S4A and B. The two brightest correspond to the periodicity of the planes in reciprocal space. To get a quantitative value of the lattice periodicity we calculate the power spectral densities (PSD) by WSxM (*3*), which basically performs squared fast Fourier transformation (*4*) of the grey scale surface image:

$$PSD\left(f_x, f_y\right) = \lim_{d \to \infty} \frac{1}{d^2} \left| \int_{-d/2}^{d/2} \int_{-d/2}^{d/2} S(x,y) \times \exp[-2\pi i(xf_x + yf_y)] \, dx dy \right|^2$$

Here $S(x,y)$ represents the 2 dimensional matrix of the color values of the TEM image, $f_x$ and $f_y$ are the spatial frequencies in $x$ and $y$ direction, respectively. The WSxM software also directly provides 1 dimensional PSD for digitized data in the vector direction where the intensities are highest, which is shown in Fig. S4C. The inverse of the peak position in this PSD spectra is the lattice constant of PZT, which is 4.23 Å for the bottom portion and 4.13 Å for the top portion. The difference between lattice constants near two surfaces indicates that during growth the PZT near LSMO was more compressively strained than PZT near top surface.

## 3. d$_{33}$ Coefficient of PZT

All the d$_{33}$ coefficient measurements are done in a local configuration, i.e. without any global top electrode. The scanning probe microscopy tip is used to apply the dc+ac electric field between PZT and the grounded substrate (Si or LSMO). Fig. S5A-C show the |d$_{33}$|, phase and |d$_{33}$|cosθ loops of PLD grown, 30 nm thick PZT film on 20 nm LSMO coated STO substrate. For a comparison, PZT from this source substrate is transferred onto another similar STO/LSMO20 substrate and the d$_{33}$ loops are measured by exactly same tip as that is used to measure the source film. Fig. S5Dd-F show the d$_{33}$ loops of the transferred 30 nm thick PZT. An interesting point to note is that the coercive field is significantly increased from 0.5 MV/cm to 1.1 MV/cm after transferring. We have also successfully transferred very thin (15 nm) PZT on LSMO substrate. The d$_{33}$ loops of the transferred thin PZT are shown in Fig. S5G-I. The coercive field of the transferred 15 nm thick flake is 1.85 MV/cm. The enhancement of coercive field is not observed for the thick PZT (100 nm) samples, as shown in Fig. S5J. However, all of the d$_{33}$-

V loops are found to be asymmetric with respect to V=0 axis, which is presumably due to the work function difference between the dissimilar contact materials. A comparison between the coercive fields extracted from $d_{33}$ measurements of the as-grown and transferred PZT samples are shown in Fig. S5K.

## 4. Capacitance and polarization characteristics of transferred PZT on different substrates

To measure the C-E and P-E characteristics, $20 \times 20$ µm² sized Au pads are fabricated on top of the PZT flakes by using shadow mask and e-beam evaporator. The positive terminal is put on top of the Au pad and the conducting substrate (highly doped Si or LSMO) serves as the ground terminal. The Si substrate is p+ doped with doping concentration of $2 \times 10^{19}$/cm³. Agilent B1500 and Radiant ferroelectric tester are used for C-E and P-E measurements, respectively. For capacitance measurement, an ac signal with 200 mV p-p amplitude and 100 KHz frequency is superimposed with the dc bias. The high frequency reduces the effect of the defects and trap charges. Fig. S6A-C show the C-E of the as-grown film grown on STO/LSMO20, transferred PZT on STO/LSMO20 and Si/Au10 substrate, respectively. All the figures are asymmetric with respect to E=0 axis, which is presumably due to the asymmetric electrodes, although the transferred PZT on Si/Au10 is expected to have symmetric boundary conditions and show symmetric C-E. This happens because the top contact Au is directly evaporated on PZT, but the bottom contact is is not, resulting in not exactly same boundary conditions. When both top and bottom contacts are made by directly evaporating Au on PZT, the asymmetry disappears, which is shown in Fig. S6D. In this case, before transferring on Si/Au, 2 nm Au is evaporated at the back side of PZT (while PZT is on the stamp). Fig. S6E shows the C-E of the transferred PZT on 5 nm $Al_2O_3$ coated Si. The P-E loops of these samples are shown in Fig. S6F-J. The saturation polarization value remains similar. For PZT transferred on Si/$Al_2O_3$ substrate both C-E and P-E loops become elongated in the horizontal (voltage) axis. This is due to the additional voltage drop that happens across the $Al_2O_3$ layer. The capacitance on the Si/$Al_2O_3$ is also smaller due to the same reason. Nonetheless, both P-E and C-E show excellent ferroelectric behavior.

## 5. Transient electronic response of the transferred PZT: observation of negative differential capacitance

One of the most attractive properties of the FE material is the negative differential capacitance which has huge potential for energy harvesting applications and reducing the sub-threshold swing in field effect transistors (5,6). Capacitance is defined as $C = \frac{dQ}{dV} = [d^2U/dQ^2]^{-1}$, where $V$, $U$ and $Q$ are the voltage, energy and charge, respectively. The negative capacitance can be understood from the curvature of the double well shape of the energy landscape originated from the Landau's model of ferroelectric materials (*7*). Although predicted several decades ago, direct observation of negative capacitance has been reported very recently (*8*). By following the experimental techniques stated in Ref. 8 we have been able to observe the similar phenomenon in the transferred PZT on STO/LSMO20 substrate.

Fig. S7A and S7B show the negative capacitance measurement setup and the equivalent circuit, respectively. A 50 KΩ series resistance is added to the FE capacitor and the voltages across the source Agilent 81150a ($V_S$) and FE ($V_F$) have been probed by digital oscilloscope with 100X probes. The parasitic capacitance from the measurement setup is 40 pF. The system is initialized by a +12 V pulse. Then we apply two consecutive negative pulses with 500 μs pulse width: 0 V→-12 V→0 V→-12 V→0V. During the first negative pulse the FE polarization switches and while switching, it traverses through negative capacitance branch. This is manifested by the positive slope of the $V_F$-t curve in figure S7D during the time period from -498 μs to -491 μs. In this time the accumulated charge-time curve has a negative slope, which results in $C = \frac{dQ}{dV} < 0$. Since the FE does not switch during the second negative pulse, we do not observe the negative capacitance and positive sloped branch in $V_F$-t.

**6. Single crystal ferroelectric gated field effect transistor on Si**

*6.1 Interface*

For field effect transistor it is imperative that the interface with the gate oxide is as much defect-free as possible. Measurement of the frequency (f) dependent capacitance of the gate material-Si stack is a vital tool to probe the defect states. At low frequency the interface trap states contribute to the total capacitance whereas at high frequency excitation their effect is negligible. We transferred PZT (100 nm) on to conductive Si, which was covered by 6 nm of thermally grown $SiO_2$. The thickness of the $SiO_2$ layer is confirmed by the TEM image. Figure S8A shows the C-f of the Si-$SiO_2$ only and Si-$SiO_2$-PZT stack, measured at zero dc bias and by 200 mV p-p ac voltage. The slopes of the C-f curves are very similar, indicating that the transfer of PZT does not alter the defect states at Si/$SiO_2$ interface. It also suggests that PZT does not bring in extra trapped charge to the system. The admittance angles, derived from the conductance and capacitance are close to 90° for both cases. Similar behavior has been observed for the transferred PZT on Si/$Al_2O_3$ (5 nm) and $Al_2O_3$ capacitors on Si (figure S9A and S9B).

*6.2 Device fabrication*

N-channel FD-SOI (fully depleted silicon on insulator) MOSFETs were fabricated on lightly doped p-type (~$10^{16}$/$cm^3$) silicon-on-insulator (SOI) wafers with buried oxide (BOX) thickness of ~200nm.

Thermal oxidation is used to thin the SOI layer down to ~100nm. The active area is patterned by optical lithography followed by dry etching. A sacrificial thermal oxide (~3nm) is grown to reduce the etch damage. After using diluted HF to remove the sacrificial oxide, a 3nm gate thermal oxide was grown immediately. The channel region is patterned by optical lithography, and then ion implantation is performed to dope the source/drain regions n-type ($5\times10^{15}$ $As^+/cm^2$ at 80keV, 7° tilt). Rapid thermal annealing (20s @ 900 °C in $N_2$) is used to activate the dopants. Forming gas annealing (25 minutes @ 350 °C) was performed to improve $Si/SiO_2$ interface properties. Then PZT with the stamp is transferred on the channel region. The PMMA is washed away by acetone, leaving PZT on the channel region. Subsequently gate electrode patterning by optical lithography and Au deposition by thermal evaporation were done.

Fig. S10 shows the optical image of a typical transistor. We fabricated a number of transistors and in this Supplementary section we show the results of a transistor different than the main text's one. Fig. S11A shows the $I_D$-$V_G$ (top gate) at $V_G$ (back gate) =-4V at different $V_D$. Note that the ON current increases with increasing $V_D$. All the curves show anti-clock wise hysteresis. The shape of the loop can be understood by noting the following: When the transistor is ON, the ferroelectric is essentially between two metal plates. Therefore, the switching of the polarization happens just it does in a usual capacitor structure and a reasonably sharp transition is seen when the transistor channel turns OFF. On the other hand, starting from OFF, the ferroelectric is between a metal plate (top electrode) and an insulator. The insulator slowly turns into a metal as the voltage across it is increased. Therefore, the actual voltage drop across the ferroelectric varies non-linearly as the total gate voltage is increased. Hence the transition from OFF to ON stretches out and shows a slower transition from OFF to ON than from ON to OFF. When top gate is grounded and only back gate voltage is swept, no hysteresis is seen in the $I_D$-$V_G$ (see Fig. S11B) indicating that the observed hysteresis for the top gate sweep indeed comes from the ferroelectric polarization.

Figure S12 shows $I_D$-$V_G$ (top gate) characteristics of a different sample where two different $I_D$-$V_G$ sweeps were performed with a time gap of 1 day. The two measurements show identical behavior, indicating that the polarization retention is robust and its effect on the Si channel is repeatable.

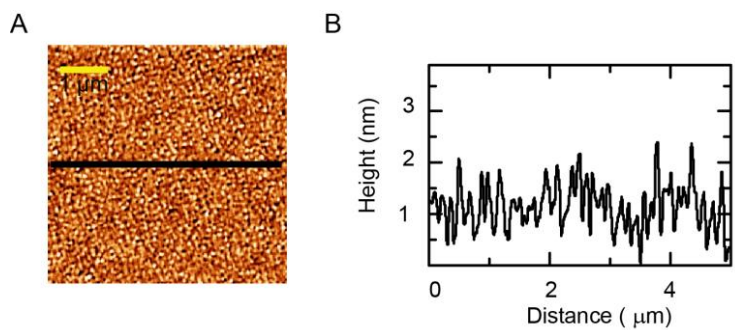

**Fig. S1. Surface morphology of the source STO/LSMO20/PZT100 film.** (**A**) Topography image. (**B**) cross-sectional height profile.

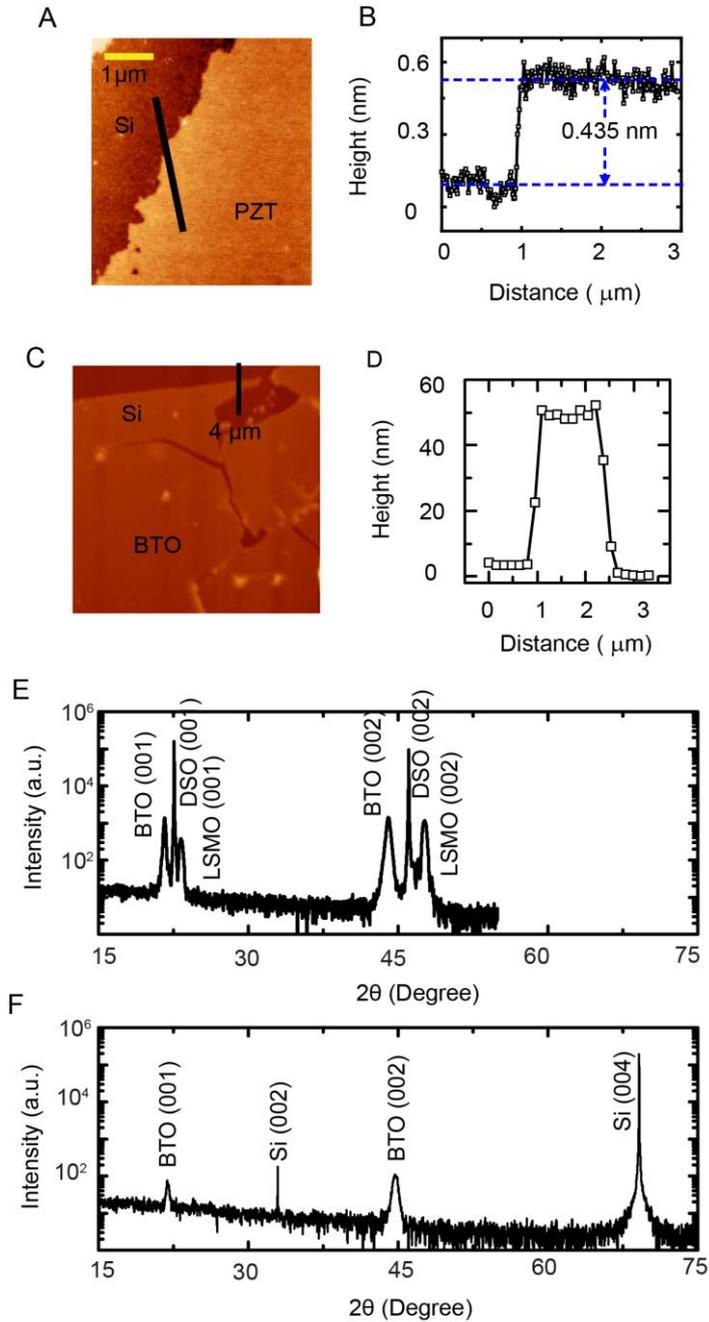

**Fig. S2. Structural characterization of ultrathin PZT and BTO**. Surface morphology and cross-sectional height profile of the transferred 1 unit cell thick PZT on Si (**A and B**) and transferred 55 nm thick BTO on Si (**C and D**). The black line indicates the place where cross-sectional height profile is taken. (**E** and **F**) XRD θ-2θ scan of BTO film on DSO/LSMO substrate and transferred BTO on Si, showing the crystallinity of BTO remains intact after transfer process.

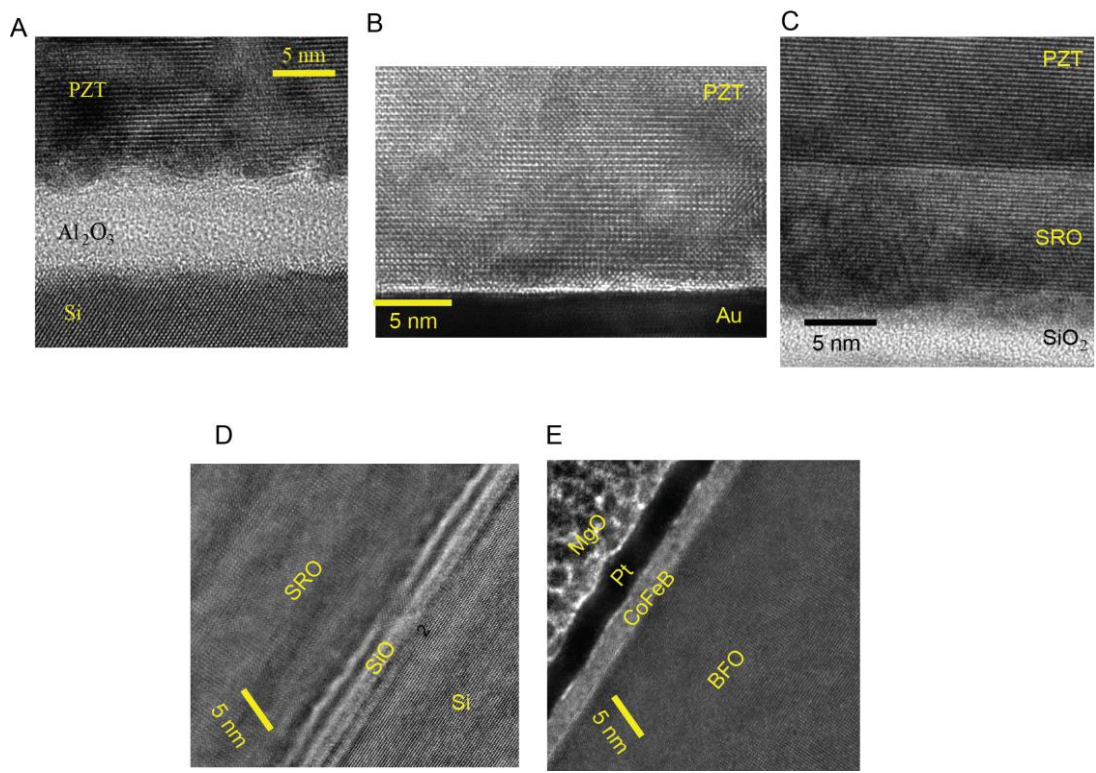

**Fig. S3. TEM images.** Transferred PZT on Si/Al$_2$O$_3$ (**A**) and Si/Au surface (**B**). (**C**) Transferred SRO/PZT/SRO heterostructure on Si/SiO$_2$. (**D** and **E**), Transferred CTO/STO superlattices on Si/SiO$_2$. (**F** and **G**) Bottom and top part of the transferred SRO/BFO/CoFeB/Pt on Si/SiO$_2$. The TEM images confirm that wide range of single crystal oxide materials can be pristinely transferred onto any type of substrate.

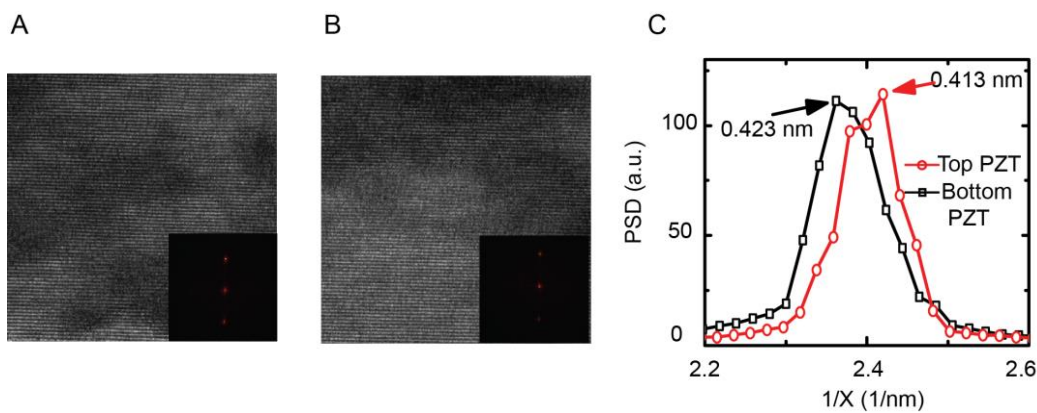

**Fig. S4. Lattice constant extraction from the Fourier transformation of the TEM images of transferred PZT near top and bottom interface.** (**A** and **B**) TEM images of transferred PZT's bottom and top portion, respectively. The insets show the color coded Fourier transformed images, where the distance of the brightest spot from the center is an inverse measure of the lattice constant. (**C**) Power spectral densities of the FT images, where the peak position corresponds to the most dominant frequency of the features present in the TEM images and its inverse is a direct measure of the average lattice constant. Clearly, the bottom portion of the transferred PZT is more compressively strained than the top portion.

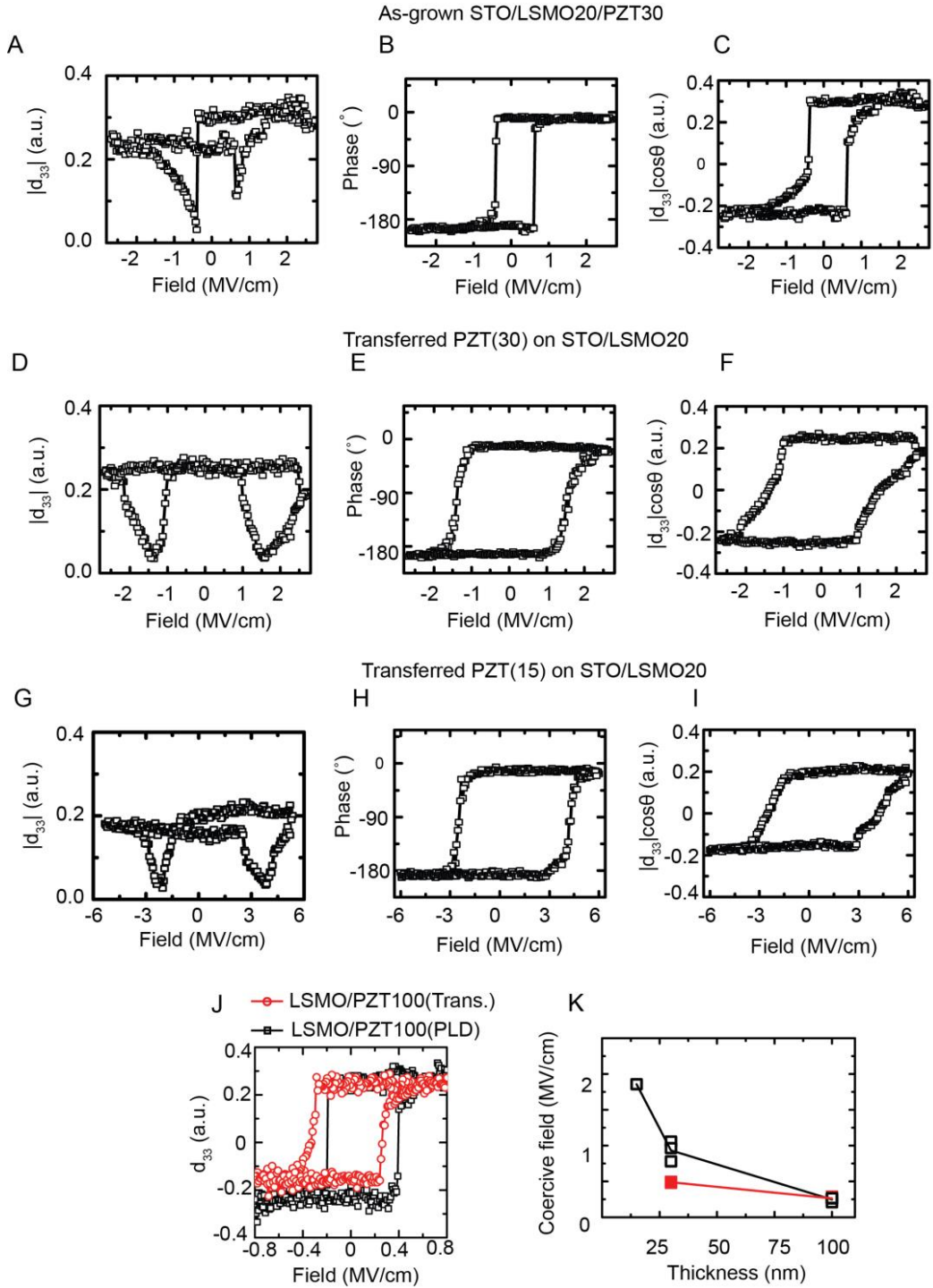

**Fig. S5. Thickness dependence of the $d_{33}$ coefficients of as-grown PZT and transferred PZT on same substrate.** Amplitude, phase and $|d_{33}|\cos\theta$ of as-grown STO/LSMO/PZT30 (**A** to **C**), transferred 30 nm thick PZT on STO/LSMO substrate (**D** to **F**) and transferred 15 nm thick PZT on STO/LSMO substrate (**G** to **I**). (**J**) Comparison between d33 coefficients of PLD-grown, 100 nm thick source PZT film on STO-LSMO substrate and transferred PZT on same substrate. $d_{33}$ Coefficients are similar in magnitude for both transferred and as-grown PZT, proving that transfer process does not significantly modify the electro-mechanical properties of PZT. (**k**) Thickness dependence of the coercive fields of as-grown and transferred PZT.

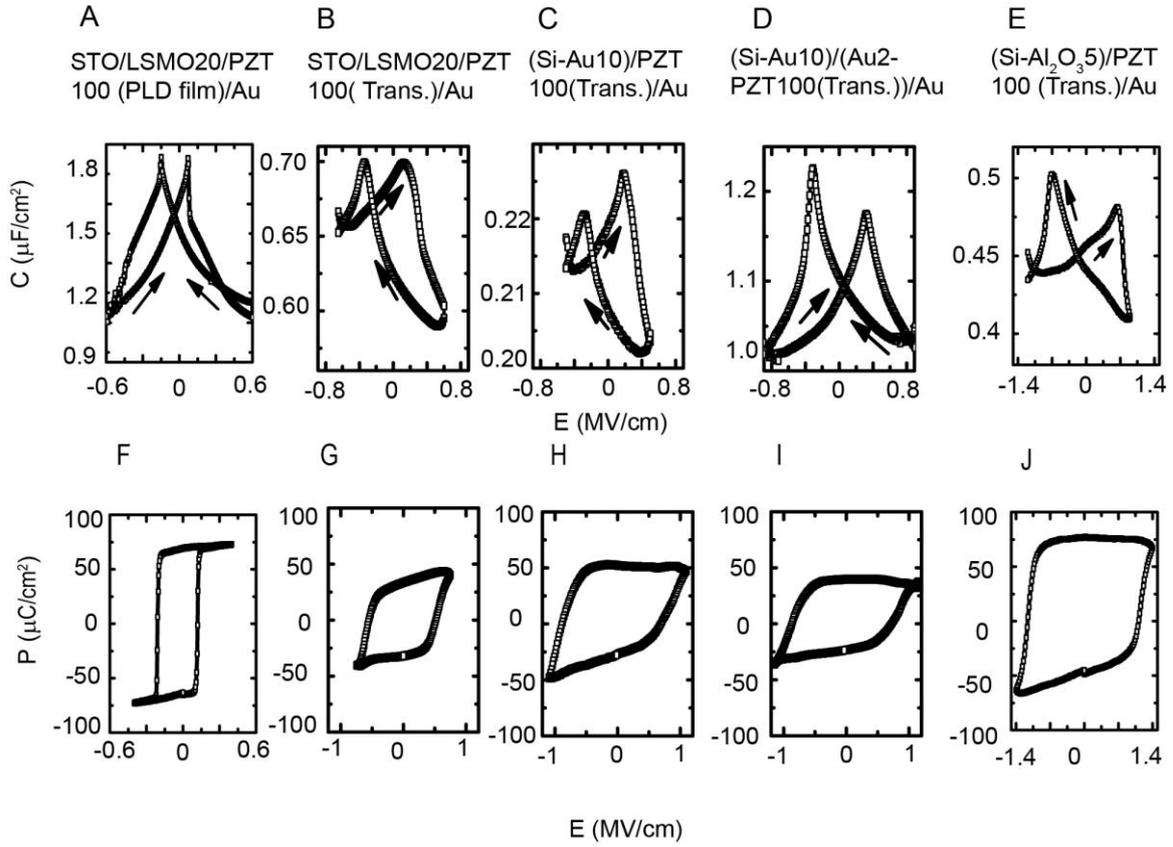

**Fig. S6. C-E and P-E of as-grown and transferred PZT (100 nm thick) on different substrates.**
C-E loops of (**A**) the as-grown (by PLD) PZT film on STO/LSMO20, (**B**) the transferred PZT on STO/LSMO20, (**C**) the transferred PZT on Si/Au10, (**D**) the transferred PZT (with 2 nm evaporated Au at the bottom surface) on Si/Au10 and **e,** the transferred PZT on Si/Al$_2$O$_3$ (5 nm), respectively. (**F** to **J**) P-E loops of the above mentioned samples.

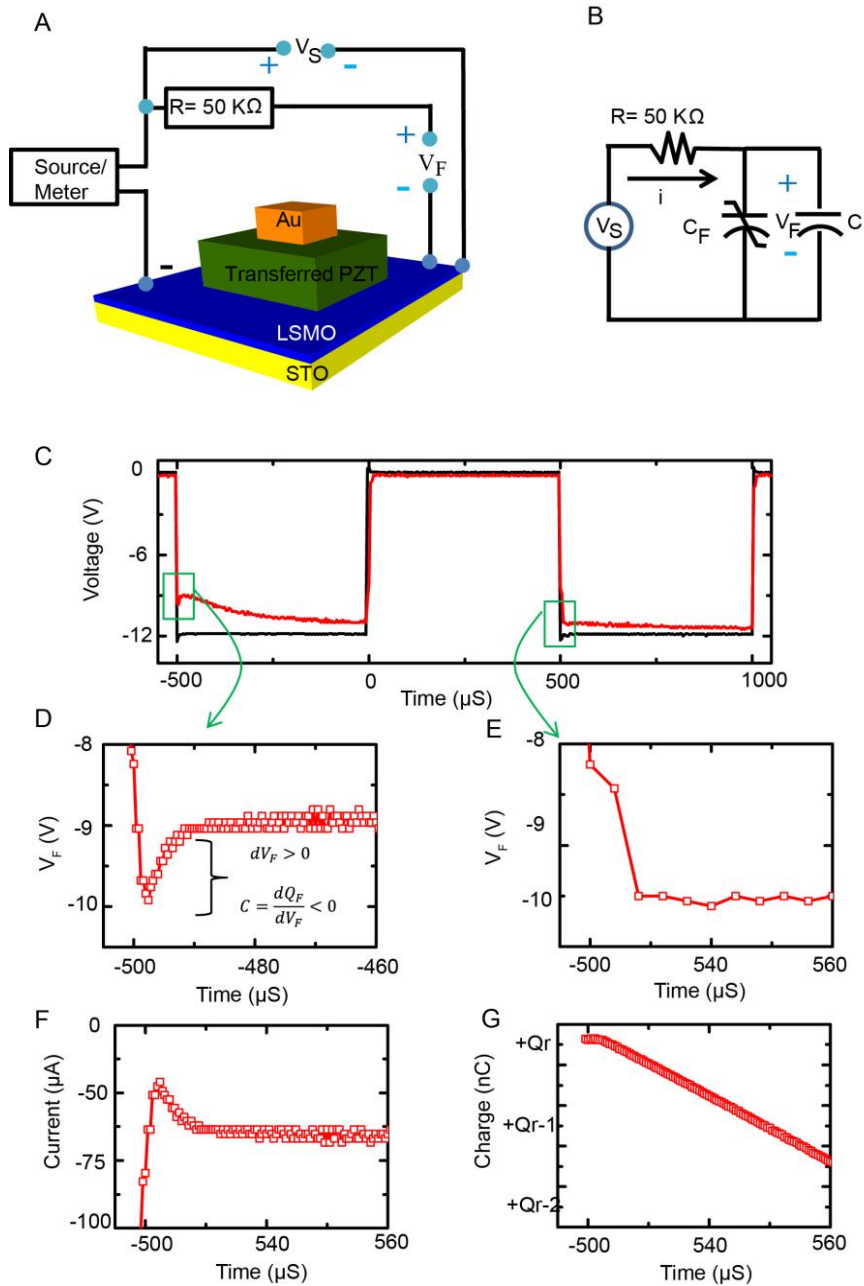

**Fig. S7. Negative capacitance in transferred PZT on STO/LSMO substrate.** (**A**) Schematic diagram of the transient characteristics measurement setup. (**B**) Equivalent circuit diagram. (**C**) Voltage waveform across the power source ($V_S$) and the FE capacitor ($V_F$). (**D** to **E**) Zoomed-in view of the transient response across FE during first and second pulse, respectively. (**F**) Charging current through FE. (**G**) Stored charge in FE.

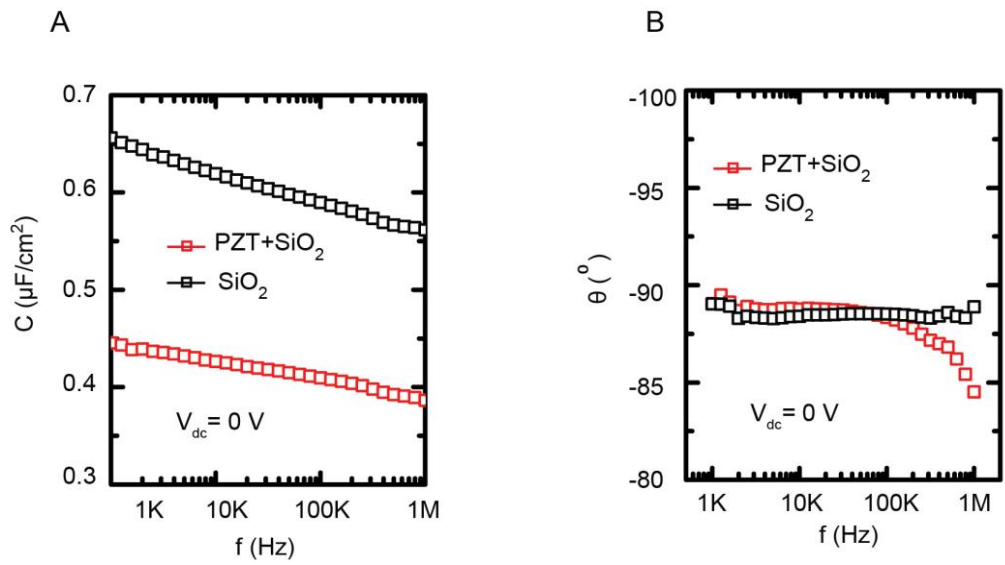

**Fig. S8. C-f data**. (A) Frequency dependent capacitance and (**B**) admittance angles of Si-SiO$_2$-transferred PZT and Si-SiO$_2$ capacitors.

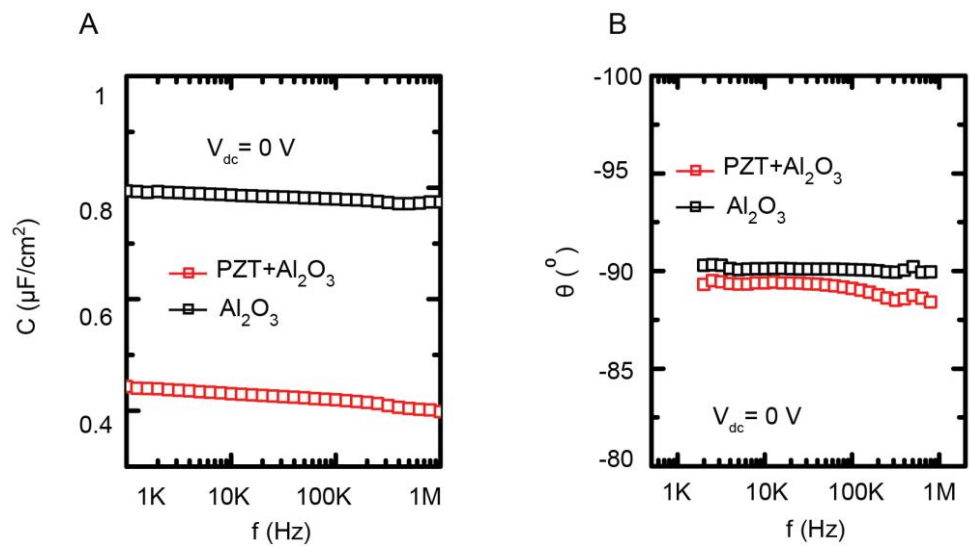

**Fig. S9. C-f data**. (**A**) Frequency dependent capacitance and (**B**) admittance angles of Si-Al$_2$O$_3$-transferred PZT and Si-Al$_2$O$_3$ samples.

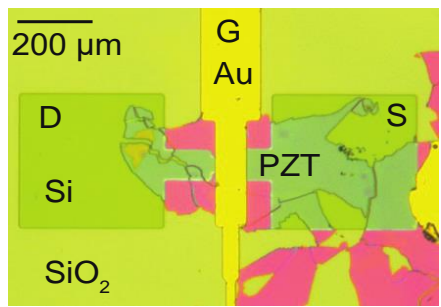

**Fig. S10. SOI Transistor.** Optical micrograph of the fabricated transistor.

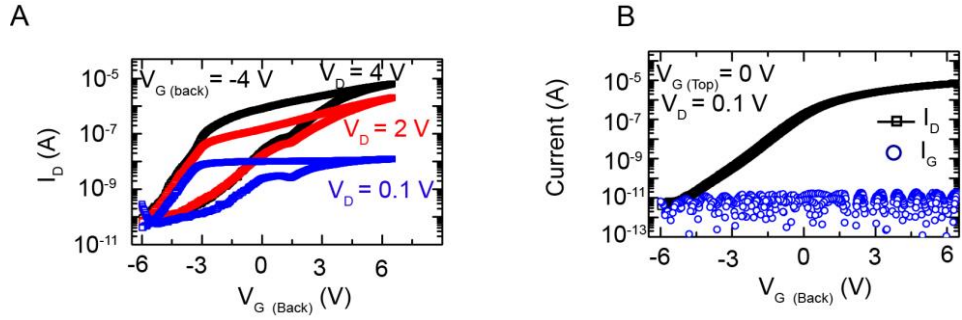

**Fig. S11. $I_D$-$V_G$ characteristics of FEFET.** (**A**) $I_D$-$V_G$ (top gate) characteristics of the ferroelectric PZT gated transistor at $V_G$ (back gate)= -4 V for different $V_D$. (**B**), $I_D$-$V_G$ (back gate) characteristics of the transistor.

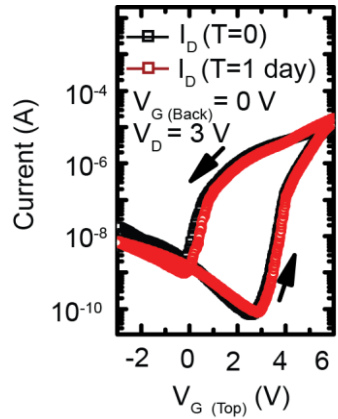

**Fig. S12. Retention.** $I_D$-$V_G$ (top gate) characteristics of a FEFET (channel length and width 5 μm) measured at a time gap of 1 day.

**Supplementary References:**

1. L. Jiao et al., Creation of nanostructures with Poly (methyl methacrylate)-mediated nanotransfer printing. J. Am. Chem. Soc. 130, 12612 (2008).

2. G. Bridoux, et al., An alternative route towards micro- and nano-patterning of oxide films. Nanotechnology 23, 085302 (2012).

3. I. Horcas et al., WSXM: A software for scanning probe microscopy and a tool for nanotechnology. Rev. Sci. Instrum. 78, 013705 (2007).

4. A. Duparré et al., Surface characterization techniques for determining the root-mean-square roughness and power spectral densities of optical components. Applied Optics 41, 154-171 (2002).

5. S. Salahuddin, S. Datta, Use of negative capacitance to provide voltage amplification for low power nanoscale devices. Nano Lett. 8, 405-410 (2008).

6. V. V. Zhirnov, R. K. Cavin, Negative capacitance to the rescue? Nat. Nanotech. 3, 77–78 (2008).

7. M. E. Lines, A. M. Glass, Principles and Applications of Ferroelectrics and Related Materials (Clarendon, 2001)

8. A. I. Khan et al., Negative capacitance in a ferroelectric capacitor. Nat. Mat. 14, 182-185 (2015).